\input harvmac
%\draftmode
\def \td {\tilde}
\def \const {{\rm const}}

\def\a{\alpha}
\def\b{\beta}
\def\g{\gamma}

\def\r{\rho}

\def\te{\theta}

\def\p{\phi}

\def \w {{\rm w}}
\def \om {\omega}

\def \r {\rho}\def \g {\gamma}  
\def \ov {\over }\def \b {\beta}
\def \P {\Phi}  \def \const {{\rm const}}

\def \ha {{1 \ov 2}}

\def \F {\varphi}

\def\a{\alpha}
\def\b{\beta}
\def\N{{\cal{N}}}
\def\g{{\gamma}}

%%%%%%%%%%%%%%%%%%%%%%%%%%%%%%%%%%%

\def  \rs { $R \times S^3$\ }
\def  \sf { $S^4$\ }
 
\def \om {{\rm w}}

\lref\WS{
L.~Susskind and  E.~Witten,
``The Holographic Bound in Anti-de Sitter Space,''
hep-th/9805114.
%%CITATION = HEP-TH 0105114;%%
A.~W.~Peet and J.~Polchinski,
``UV/IR relations in AdS dynamics,''
{\it Phys.\ Rev.\ D } {\bf 59}, 065011 (1999), 
hep-th/9809022.
%%CITATION = HEP-TH 9809022;%%
}

\lref\ghkt{
S.~S.~Gubser, C.~P.~Herzog, I.~R.~Klebanov and A.~A.~Tseytlin,
``Restoration of chiral symmetry: A supergravity perspective,''
{\it JHEP } {\bf 0105}, 028 (2001), 
hep-th/0102172.
%%CITATION = HEP-TH 0102172;%%
}

\lref\KT{
I.~R.~Klebanov and A.~A.~Tseytlin,
``Gravity Duals of Supersymmetric $SU(N) \times
SU(N+M)$ Gauge Theories,''
{\it Nucl. Phys.} {\bf B578} (2000) 123,
hep-th/0002159.
%%CITATION = HEP-TH 0002159;%%
}

\lref\Buch{A. Buchel, ``Finite temperature resolution
of the 
Klebanov--Tseytlin singularity,'' 
{\it Nucl.\ Phys.\ B }
{\bf 600}, 219 (2001),  hep-th/0011146.
%%CITATION = HEP-TH 0011146;%%
}

\lref\KS{I.~R.~Klebanov and M.~J.~Strassler, 
``Supergravity and a Confining Gauge Theory:  
Duality Cascades and $\chi$SB-Resolution of Naked
Singularities,''
{\it JHEP} {\bf 0008} (2000) 052,
hep-th/0007191.
%%CITATION = HEP-TH 0007191;%%
}

\lref\PTs{L.~A.~Pando Zayas and A.~A.~Tseytlin,
``3-branes on resolved conifold,''
{\it JHEP} {\bf 0011} (2000) 028,
hep-th/0010088.
%%CITATION = HEP-TH 0010088;%%
}

\lref\Witt{E.~Witten,
``Anti-de Sitter space, thermal phase transition, and
confinement in gauge
theories,''
{\it Adv.~Theor.~Math.~Phys.} {\bf 2} (1998) 505,
hep-th/9803131.
%%CITATION = HEP-TH 9803131;%%
}

\lref\wittH{E.~Witten,
``Anti De Sitter Space And Holography,''
{\it Adv.~Theor.~Math.~Phys.} {\bf 2} (1998) 253,
hep-th/9802150.
%%CITATION = HEP-TH 9802150;%%
}

\Title{\vbox
{\baselineskip 10pt
%\hbox{NSF-ITP-01-}
 \hbox{OHSTPY-HEP-T-01-028} }}
{\vbox{\vskip -30 true pt\centerline {
Curved space resolution of singularity}
\medskip
\centerline { of  fractional 
D3-branes on conifold
 }
\medskip
\vskip4pt }}
\vskip -20 true pt
\centerline{A. Buchel$^{1}$ and A. A.
Tseytlin$^{2}$\footnote{$^*$} {Also at
Imperial College,   London  and 
 Lebedev 
Physics Institute, Moscow. }
 }
\smallskip\smallskip
\centerline{$^{1}$ \it Institute for Theoretical
Physics,
University of California, Santa Barbara, CA
93106, USA}
\centerline{$^{2}$ \it  Department of Physics, 
The Ohio State University,  
Columbus, OH 43210, USA }

\bigskip\bigskip
\centerline {\bf Abstract}
\baselineskip12pt
\noindent
\medskip
We construct a supergravity dual to the 
 cascading $SU(N+M)\times SU(N)$ supersymmetric 
gauge theory (related to fractional D3-branes on conifold
according to Klebanov et al) in the case when the 
3-space  is compactified 
on $S^3$ and in  the  phase with unbroken chiral symmetry. 
The size of $S^3$ serves as an infrared cutoff
on the gauge theory dynamics. For a  sufficiently 
large $S^3$ the dual  supergravity background
 is expected to be nonsingular.     
We demonstrate that this is indeed the case: we  find  a smooth 
type IIB supergravity solution  using a perturbation 
theory that is valid when the radius of $S^3$ is large. 
We  consider also  the 
 case with  the euclidean world-volume  being  $S^4$
instead of  $R\times S^3$,   where the supergravity solution
is again  found to be regular.
This ``curved space'' resolution of the  singularity 
of the fractional D3-branes  on conifold 
solution 
is analogous to the one in the non-extremal (finite temperature)
case discussed in our previous work.

\bigskip

%%%%%%%%%%%%%%%%%%%%%%%%%%%%%%%%%%%%%%%%%%%%%%%%%%%%%%%%%
\Date{11/01}

%%%%%%%%%%%%%%%%%%%%%%%%%%%%%%%%%%%%%%%%%%%%%%%%%%%%%%%%%%%%%%%%%%%
\noblackbox 
\baselineskip 16pt plus 2pt minus 2pt

\lref\fiv{
A.~Buchel, C.~P.~Herzog, I.~R.~Klebanov, L.~Pando Zayas and
A.~A.~Tseytlin,
``Non-extremal gravity duals for fractional D3-branes on the
conifold,''
{\it JHEP} {\bf 0104}, 033 (2001), 
hep-th/0102105.
%%CITATION = HEP-TH 0102105;%%
}

\lref\agmoo{O.~Aharony, S.~S.~Gubser, J.~Maldacena, H.~Ooguri and Y.~Oz,
``Large N field theories, string theory and gravity,''
{\it Phys.\ Rept.\ }  {\bf 323}, 183 (2000), 
hep-th/9905111.
%%CITATION = HEP-TH 9905111;%%
I.~R.~Klebanov,
``Introduction to the AdS/CFT correspondence,''
hep-th/0009139.
%%CITATION = HEP-TH 0009139;%%
}

\lref\kw{I.~R.~Klebanov and E.~Witten,
``Superconformal field theory on threebranes at a Calabi-Yau 
singularity,''
{\it Nucl.\ Phys. B}\  {\bf 536}, 199 (1998), 
hep-th/9807080.
%%CITATION = HEP-TH 9807080;%%
}

\lref\kn{I.~R.~Klebanov and N.~A.~Nekrasov,
``Gravity duals of fractional branes and logarithmic RG flow,''
{\it Nucl.\ Phys.\ B}  {\bf 574}, 263 (2000), 
hep-th/9911096.
%%CITATION = HEP-TH 9911096;%%
}
\lref\ps{J.~Polchinski and M.~J.~Strassler,
``The string dual of a confining four-dimensional gauge theory,''
hep-th/0003136.
%%CITATION = HEP-TH 0003136;%%
}
\lref\mn{
J.~M.~Maldacena and C.~Nunez,
``Towards the large N limit of pure N = 1 super Yang Mills,''
{\it Phys.\ Rev.\ Lett.\ } {\bf 86}, 588 (2001), 
hep-th/0008001.
%%CITATION = HEP-TH 0008001;%%
}
\lref\bufr{A.~Buchel and A.~R.~Frey,
``Comments on supergravity dual of pure N=1 super Yang Mills theory with
 unbroken chiral symmetry,''
{\it Phys.\ Rev.\ D} {\bf 64}, 064007 (2001), 
hep-th/0103022.
%%CITATION = HEP-TH 0103022;%%
}

\lref\gtv{
S.~S.~Gubser, A.~A.~Tseytlin and M.~S.~Volkov,
``Non-Abelian 4-d black holes, wrapped 5-branes, and their dual 
descriptions,''
{\it JHEP} {\bf 0109}, 017 (2001)
hep-th/0108205.
%%CITATION = HEP-TH 0108205;%%
}

\lref\hkq{
C.~P.~Herzog, I.~R.~Klebanov and P.~Ouyang,
``Remarks on the warped deformed conifold,''
hep-th/0108101.
%%CITATION = HEP-TH 0108101;%%
}

\lref\war{
A.~Khavaev and N.~P.~Warner,
``An N = 1 supersymmetric Coulomb flow in IIB supergravity,''
hep-th/0106032.
%%CITATION = HEP-TH 0106032;%%
}

\lref\Gubser{
 S.~S.~Gubser,
``Curvature Singularities: the Good, the Bad, and the Naked,''
hep-th/0002160.
%%CITATION = HEP-TH 0002160;%%
}

\lref\PW{
K.~Pilch and  N.~P.~Warner,
``N=2 Supersymmetric RG Flows and the IIB Dilaton,''
{\it Nucl.\ Phys.\ B}   {\bf 594}, 209 (2001), hep-th/0004063.
%%CITATION = HEP-TH 0004063;%%
}

\lref\bpp{
A.~Buchel, A.~W.~Peet and J.~Polchinski, 
``Gauge Dual and Noncommutative Extension of an N=2
 Supergravity Solution,''
 {\it Phys.\ Rev.\ D} {\bf 63}, 044009 (2001), 
hep-th/0008076.
%%CITATION = HEP-TH 0008076;%%
}

\lref\j{
N.~Evans, C.~V.~Johnson and  M.~Petrini, 
`The Enhancon and N=2 Gauge Theory/Gravity RG Flows,''
{\it JHEP}   {\bf 0010}, 022 (2000),  hep-th/0008081.
%%CITATION = HEP-TH 0008081;%%
}

%%%%%%%%%%%%%%%%%%%%%%%%%%%%%%%%%
\newsec{Introduction}
%%%%%%%%%%%%%%%%%%%%%%%%%%%%%%%%%%%
Gauge theory -- gravity duality\foot{For  reviews
and references see, e.g.,  \agmoo.} relates a gauge theory on the world
volume of a 
large number of D-branes to  supergravity 
backgrounds where the 
branes are replaced by the corresponding fluxes.
In a particular  realization of this duality, Klebanov and Witten (KW)
\kw\ 
considered $N$ regular D3-branes placed at a conical singularity    
in type IIB string theory. At small 't Hooft coupling 
$g_s N\ll 1$, the system is best described by open strings and 
realizes $SU(N)\times SU(N)$ $\N=1$ supersymmetric 
gauge theory with two pairs 
of chiral multiplets $A_i,\ B_j$ and a quartic  superpotential
at an infrared superconformal fixed point. 
In the limit of strong 't  Hooft coupling this gauge theory 
is best described by type IIB supergravity compactified on 
$AdS_5\times T^{1,1}$, $T^{1,1}=(SU(2)\times SU(2))/U(1)$,
with $N$ units of the RR 5-form flux through the $T^{1,1}$.
 If this 
is a genuine equivalence, then  phenomena observed on the 
gauge theory side should have  dual description  in string theory
on  $AdS_5\times
T^{1,1}$.
In particular,  {\it any} deformation of
the gauge theory visible in the 
large $N$ limit should have a counterpart 
in the dual gravitational
description, and vice versa.

 Certain deformations, trivial on 
the  gravity side,  may  have  highly nontrivial analogs  in the 
gauge theory dynamics.  
For example, 
the  presence of the $AdS_5$ factor in the KW geometry is 
a reflection 
of the conformal symmetry of the dual gauge theory.  In the  Poincare 
coordinates in  $AdS_5$, 
 its  boundary, and thus the space-time 
where the gauge theory is formulated, is $R^{1,3}$. 
In the global 
parameterization of  $AdS_5$ the boundary is $R\times S^3$.
Such  gravitational 
background should correspond to the 
the  superconformal 
KW gauge theory defined on  $R\times S^3$. 
{}From the supergravity perspective,   
going from the  Poincare  to the global coordinates is a simple 
local coordinate transformation.  However, 
on  the gauge 
theory side, 
this ``deformation'' drastically modifies the dynamics. 
Defined on a  round 3-sphere the gauge theory will 
have no zero  modes:\foot{For the scalars, this follows from  their
coupling 
to the scalar 
curvature,  required by the conformal invariance.
}
it will have 
a mass gap in the spectrum of  order of  inverse 
radius of $S^3$. 
The modification of the spectrum of the theory
substantially modifies its thermodynamics. As in a similar system 
studied in \Witt, we expect a  thermal phase transition in the 
$S^3$-compactified KW model, which,   on the gravity 
side, should   map into 
 the Hawking-Page phase transition.
  We would like to emphasize 
that such a phase transition should occur only for the 
gauge theory defined on a curved space like  $S^3$.

It is  not known how to translate a generic 
gauge theory deformation into
the  dual supergravity description.
 For the   particular deformations for which the dictionary
is known, one typically encounters a naked singularity in the 
corresponding  deformed geometry. 
Consider, for example,  wrapping    
$M$ D5-branes 
on the collapsed 2-cycle of the conifold,  in addition 
to $N$ D3-branes put at  its apex \kn. On the gauge theory side 
this deformation corresponds to changing the gauge group to 
$SU(N+M)\times SU(N)$ with the same set of chiral multiplets
and the superpotential as in the $M=0$ case. The 
dual supergravity  background found  in \KT\ 
was shown to have a naked singularity. Another example 
with a naked singularity in the bulk
is provided by a large number on NS5-branes wrapping a 2-cycle 
of the resolved conifold in type IIB string theory \mn.
The field-theory 
 dual of this system can be interpreted as 
a  compactification (in our language -- a  deformation) of 
the little string theory on  $S^2$.
 Yet another, probably the simplest, 
example of  generation of  IR  singularity 
is a mass deformation of the $\N=4$ $SU(N)$ SYM theory dual to 
$AdS_5\times S^5$ compactification of type IIB string theory.
Turning on a mass deformation on the gauge theory side
translates into turning on 3-form fluxes on the gravity side 
\refs{\wittH,\ps,\war}.
At the linearized level, the fluxes diverge in the bulk, leading to
a naked singularity. 

A common feature of the discussed singularities 
is that they are produced by a well-defined deformation in the dual 
gauge-theory 
 system. On the gravity side they occur in the bulk 
 (as opposed to the boundary) 
 of the
geometry,
which,  according to the familiar UV/IR correspondence \WS\
expected in 
gauge--gravity 
duals,
 should 
 reflect the IR physics of the gauge theory. 
If we can  resolve
 the IR singularity induced by 
the deformation on the  gauge theory side, 
then the translation of the
resolution mechanism 
to the gravity side should cure  the  naked
singularity  there as well.

 This philosophy is rooted in the 
belief that there is a
  genuine equivalence between the two dual descriptions, 
    and 
  it was recently 
successfully  applied, in particular,  in 
refs. \refs{\ps,\KS,\mn} and     in  
\refs{\Buch,\fiv,\ghkt,\bufr,\gtv}. 
These two groups of 
papers   differ in the type of  mechanism
  used for the  singularity 
resolution. In the former case,
 the singularity in the deformed 
gauge theory is resolved by  non-perturbative phenomena, 
intrinsic to  gauge theory, namely,  the  confinement and 
the chiral symmetry breaking.
 The resolution of the 
singularity proposed in  the second group of papers is 
 extrinsic to  gauge theory:  
one puts  the system at (sufficiently high)
  finite temperature.

In this paper we 
 propose a more unified perspective on the issue of 
singularity resolution in 
gauge--gravity duals, and present a new specific 
example of the resolution  mechanism.
Although  we shall concentrate on the case of 
the fractional D3-branes on conifold  geometry \KT\
(KT background for short), 
we believe that our 
 approach is generic and should be applicable to other cases 
as well. 

An 
overview of the singularity resolution approaches
 given above  underscores 
the similarity in all resolution mechanisms. As we 
have emphasized,
in all cases the singularity is an IR phenomenon  
when viewed from the 
gauge theory perspective. 
Then a  natural  way
to 
resolve the singularity is to disallow the  gauge theory to 
access low-energy states. 
This can be achieved as a result of a dynamical gauge-theory 
effect  (generation of a mass gap in the spectrum 
due to confinement as in \refs{\ps,\KS,\mn}) or by introducing an 
IR cutoff ``by hand`` (turning on a finite
 temperature\foot{The
proposal to use 
a finite temperature as an IR cutoff to 
cloak naked singularities 
in five dimensional gravity coupled to scalars was 
put forward in \Gubser .}  
as in \refs{\Buch,\fiv,\ghkt,\bufr,\gtv}).  It is clear 
from this perspective  that there should be many  other 
ways  to resolve the singularity: all one has to do is to 
introduce an IR cutoff on the field theory side 
and to understand  what that cutoff translates into  on 
 the gravity side of the duality. 
 The corresponding supergravity background 
 should contain an extra scale 
 (the deformed conifold scale in \KS,  or the  non-extremality 
 parameter in \refs{\Buch,\fiv,\ghkt}, 
 or the  curvature of the ``longitudinal'' space in the examples
 considered  below).
 
One possibility  to  introduce an 
IR cutoff is by  ``compactifying''  the  space on 
which the   gauge theory is defined.
As a  specific realization 
of this proposal we shall consider the resolution of the  
singularity of the KT background
by defining  the dual gauge theory on  $R\times S^3$
instead of 4-d Minkowski space. 
The space  compactification  should provide
 an IR cutoff, and so for 
sufficiently large radius of the 3-sphere we 
should expect the  restoration 
of  chiral symmetry in the dual field theory,
 and thus\foot{The singularity
of the  KT  background    
is related \KS\ to  the chiral symmetry breaking  in 
the dual field theory.
This symmetry (reflected in the
 $U(1)$  fiber symmetry of $T^{1,1}$) will be 
 present in the generalized KT background we will 
 construct.}        a  smooth 
 dual supergravity background.

It should be  emphasized that not 
all of space compactifications (that provide an IR cutoff) 
may  resolve
the singularity of the supergravity dual. 
For example, compactifying the $SU(N+M)\times SU(N)$ 
gauge theory   on a   3-torus $T^3$
will  not resolve the singularity.\foot{The gravity 
dual will be the 
original 
KT solution \KT\ with the spatial coordinates of the D3-brane
world-volume 
periodically identified.}
We expect 
 that a  ``good'' (singularity-resolving)
 compactification is the one 
that lifts the zero modes of all of the gauge-theory 
 fields, i.e. gauge bosons, fermions and scalars. 
Let the space  on which  the gauge theory is defined 
be compactified on  a 
$d$-dimensional manifold 
${\cal M}_d$. 
There will not be massless gauge-boson modes, provided the first
Betty 
number of ${\cal M}_d$ vanishes. 
The 
scalars will not have   zero modes
 provided they are coupled to a non-zero scalar curvature 
of  ${\cal M}_d$.
Thus the second condition is a nonvanishing Ricci scalar  of 
${\cal M}_d$. 
One is also to  make sure that there is  no fermionic 
zero modes. 
While the $S^3$ compactification 
satisfies  these conditions, the $T^3$  one  fails
to do that.

One may also consider a Euclidean version  and define the 
gauge  theory  on a curved 4-d space-time, e.g., 
$S^4$ or $K3$. Then  $S^4$ will  lead to a resolution 
of the singularity (as we shall see below), but $K3 $ will not, 
since it has $R_{mn}=0$ and thus does not lift 
the zero modes of the scalars.

Let us  comment also on a peculiar 
relation between the space on which gauge
theory is defined 
and its  counterpart  in the dual supergravity description.
On the gauge theory side we think of space-time 
 being a manifold of fixed size.
  In the context of gauge theory -- gravity  duality, 
the space-time  where the gauge theory ``lives" 
should be  identified with a  
boundary submanifold of the dual 10-d supergravity space-time. 
The size  of this submanifold  may obviously 
 dependent on other (transverse) directions.
One  example is $AdS_5 \times S^5$ 
in  global parametrization of
 the $AdS_5$, where  
the size of the spatial part of 
the boundary $S^3$ changes with the  radial 
coordinate of $AdS_5$. 
Another example  is provided by 
 the duality discussed in \mn, where
 the gauge theory arises from compactification 
of the  little string theory 
 on  $S^2$ of  fixed size.
  The size of the corresponding
2-sphere  in  the dual supergravity background 
changes logarithmically with 
the radial coordinate \mn.\foot{Related
observations can be made in  the case of  other, more
familiar,  deformations of 
gauge theory. In \ps\ the authors studied the duality 
in the context of the mass deformed $\N=4$ $SU(N)$ SYM theory. There, 
a constant mass deformation on the gauge theory side 
translated into
a variable 3-form flux in the gravity dual.}

\bigskip

The rest of the paper is organized as follows. In  section 2 we discuss 
the generalizations of the KT  ansatz for the supergravity 
background dual to the cascading gauge 
theory compactified on $R\times S^3$ and $S^4$. Following  the approach
of  \refs{\KT,\fiv,\ghkt}, in section 3 we derive 
the corresponding 
1-d effective action that generates the equations for the 
radial evolution of the  functions parametrizing 
the background metric and matter fields. 
  We then  discuss the 
simplest  supersymmetric solutions of these equations realizing 
the extremal   fractional D3-brane 
  KT background \KT\ and the $AdS_5 \times T^{1,1}$ 
 gravity dual to the 
KW gauge theory \kw\ compactified on $R\times S^3$ or $S^4$.  

We then consider the   deformations  of  $M_4=R\times S^3$ 
and $M_4=S^4$ compactifications 
of the KW model  caused by switching on 
$P\ne 0$  units of fractional 3-brane flux. 
As  in the closely related work \ghkt\ on the non-extremal
 generalization
of the KT background,  
being unable to solve the 
resulting equations exactly, 
we resort to  a  perturbation theory 
valid in the regime when the effective 
D3-brane charge (or the 5-form flux)  $K_{\star}$ is 
much larger than the fractional 3-brane charge, $K_{\star}\gg P^2$.
Physically, this approximation amounts to introducing an IR cutoff 
in the dual gauge theory at an  energy scale high enough 
to mask the low energy chiral symmetry breaking,
which is responsible for the generation of
 the KT singularity \KS.\foot{This is also the region of validity
 of the non-extremal deformation, i.e. of the 
finite temperature resolution of the KT singularity due to the chiral 
symmetry restoration studied in \ghkt.}

In section 4 we  construct a smooth supergravity solution 
interpolating 
between the $S^4$ compactification of the KW model in the IR 
and the  KT model in the UV.
 In section 5 we address the same problem  in 
a technically more challenging case of the 
$R\times S^3$ compactification of the 
KT model. 
Both examples of regular compactifications of the KT model 
provide support to  the general idea of resolving 
 naked  singularities in the bulk of  gravitational duals
  to gauge theories 
by an IR cutoff produces by a ``boundary'' space compactification.

We conclude in section 6 with comments on constructing a
gravitational dual to  mass-deformed conformally compactified 
$\N=4$ supersymmetric Yang-Mills theories.

%%%%%%%%%%%%%%%%%%%%%%%%%%%%%%%%%%%%%%%%%%%%%%%%%%%%%%%
\newsec{$R\times S^3$ and $S^4$  generalizations of the KT background }
%%%%%%%%%%%%%%%%%%%%%%%%%%%%%%%%%%%%%%%%%%%%%%%%%%

Our aim will be to explore 
 the generalization of the KT solution \KT\
 for fractional D3-brane on conifold  
 to the case when the constant radial distance slices of the 
 ``parallel'' part of the metric have geometry  \rs\ or \sf\ 
 (we shall consider the case of Euclidean signature).
 We shall argue that the corresponding solutions 
 are regular (for large enough D3-brane charge compared to 
 fractional 3-brane charge).
 
We shall start with  the same ansatz  as in \refs{\KT,\fiv}
 and simply
replace 1+3 ``longitudinal'' directions by 
$R\times S^3$ or by $S^4$. 
The treatment of the two cases will be very similar, 
and we will discuss  them in parallel.
There will be direct analogy with the non-extremal (finite
temperature) case
considered in \refs{\Buch,\fiv,\ghkt}.

As in \KT\ we will  impose the requirement
that
the background has abelian  symmetry associated with
the $U(1)$
fiber of $T^{1,1}$  as we will consider a phase where chiral
symmetry is restored.
In the case of $R\times S^3$ 
our  general ansatz for a 10-d (Euclidean-signature)
 Einstein-frame metric will involve
4 functions $x,y,z$ and $w$  of  radial coordinate $u$ \foot{This
 metric 
can  be brought into a more familiar
form
$
ds^2_{10E} =  h^{-1/2}(r)  (dM_4)^2 
+  h^{1/2}(r)   [  {\rm g}^{-1} (r) dr^2 
+ r^2  ds^2_5] \ , $
where 
$
 h=  e^{-4z- 4x} ,\    r  = e^{y + x + w 
 } ,\  g=  e^{-8x}  ,\ \ e^{10y + 2x}
du^2  = {\rm g}^{-1} (r) dr^2.$
When $w=0$ and $e^{4y}=r^4= { 1 \ov 4u} $,
the transverse 6-d space is the standard conifold
with $M_5= T^{1,1}$.  
Small  $u$  thus  corresponds to large distances in 5-d 
 and vice versa. In the $AdS_5$ region 
 large $u$ is near the origin of
$AdS_5$ space, while $u=0$ is its boundary.}
\eqn\mett{
ds^2_{10E} =  e^{2z} (dM_4)^2 
+ e^{-2z}  [e^{10y} du^2 + e^{2y} (dM_5)^2] \ ,   }
\eqn\hih{
(dM_4)^2 =  e^{-6x} dX_0^2 + e^{2x} (dS^3)^2 \ , }
\eqn\tri{
(dS^3)^2 = 
 d \a^2 + \sin^2\a \ ( d \b^2 + \sin^2 \b\  d \g^2) \ . }
Here the 3-sphere replaces the  3 ``flat'' longitudinal directions
of the  3-brane 
and  $M_5$ is a deformation of the $T^{1,1}$ metric 
\eqn\mmm{
(dM_5)^2 =  e^{ -8w}  e_{\psi}^2 +  e^{ 2w}
\big(e_{\theta_1}^2+e_{\phi_1}^2 +
e_{\theta_2}^2+e_{\phi_2}^2\big)  \ , }
$$
 e_{\psi} =  {1\ov 3} (d\psi +  \cos \theta_1 d\phi_1 
 +  \cos \theta_2 d\phi_2)  \  , 
 \quad  e_{\theta_i}={1\ov \sqrt 6} d\theta_i\ ,  \quad
  e_{\phi_i}=
{1\ov \sqrt 6} \sin\theta_id\phi_i \ .
$$
We choose the radius of $S^3$ to be 1 as it 
can be absorbed into 
a shift of $x$  (and a rescaling of $X_0$).

In the case of $S^3$ replaced by $ R^3$ (i.e. in the 
 $x\to x+ x_0, \ x_0
\to \infty
$ limit) 
this becomes the ansatz of \fiv, 
where the 
non-extremal D3-brane  case was considered.
The  extremal D3-brane on the standard conifold and the more
general fractional D3-brane   
KT solution have  $x=w=0$.
While in  \fiv\  a  non-constant function  $x(u)$($=a u$)
was reflecting  the  non-extremality of the background, in 
the present 
$R \times S^3$ case 
it  will be  non-trivial as a consequence  of 
the curvature of $S^3$.

The ansatz in the $S^4$ case is the same as \mett\ but with 
$(dM_4)^2 $  given by 
\eqn\tris{
(dM_4)^2 = (dS^4)^2 = 
 d \a^2 + \sin^2\a \ [d \b^2 + \sin^2 \b\  ( d \g^2 
+ \sin^2 
\g\  d \delta^2   ) ]  \ , } 
where  the radius of $S^4$ is again chosen to be 1. 
Here  there is no function $x$, i.e. 
the number of 
functions in the metric is the same as in the extremal case
(however, in contrast to the standard KT case, here
$w$ will,  in general,  be 
 non-trivial).

As for the matter fields, 
we  shall assume that the dilaton $\P$ may depend on $u$, and 
our ansatz for the $p$-form  fields 
(the same in the \rs and \sf cases) 
 will be   exactly  
as in the extremal KT case \KT\ and in \fiv:\foot{The reason why
the form of the ansatz is the same  is that it is formulated 
in terms of the transverse space geometry only 
(the ``parallel'' or ``electric'' part of $F_5$  is  then fixed by
the selfduality condition).}
\eqn\har{
F_3 = \   P
e_\psi \wedge
( e_{\theta_1} \wedge e_{\phi_1} - 
e_{\theta_2} \wedge e_{\phi_2})\ ,
\ \ \ \ \ \ \ 
B_2  = \    f(u) 
( e_{\theta_1} \wedge e_{\phi_1} - 
e_{\theta_2} \wedge e_{\phi_2})
 \ , 
}
\eqn\fiff{
F_5= {\cal F}+*
{\cal F}\
, \quad  \ \ \ \ {\cal F} = 
K(u) 
e_{\psi}\wedge e_{\te_1} \wedge
e_{\p_1} \wedge
e_{\te_2}\wedge e_{\p_2}\  , \ \ \ \ \ \ \
 K (u)  = Q + 2 P f (u) \ , }
where, as  in \KT, the  expression for $K$ 
follows from the Bianchi identity for the
5-form. The constants $Q$ and $P$ are proportional 
to the numbers $N$ and $M$ of standard and fractional 
D3-branes; their precise normalizations 
(see \hkq) will not be important here. 

%%%%%%%%%%%%%%%%%%%%%%%%%%%%%%%%%%%%%%%%%%%%%%%%
In what follows, we shall first derive the 
corresponding  system of type IIB supergravity
equations of motion 
 describing the 
radial  evolution of the six 
unknown functions of $u$:
$x,y,z,w,K,\P$  (five functions $y,z,w,K,\P$ in the \sf case).

We shall then discuss its 
solutions  aiming to show that 
there exists a smooth interpolation (in radial coordinate only)
 between (i)
 a non-singular 
 short-distance region where the 10-d background  is 
 approximately $AdS_5 \times T^{1,1}$ written in the 
 coordinates where the $u=\const$ slice
 is $R \times S^3$
 or $S^4$, and 
 (ii) a long-distance region  where the 10-d  background  
 approaches the KT solution. 
We shall start  with the short-distance 
($u=\infty$ or $\r=0$) region, 
i.e.    $AdS_5 \times T^{1,1}$ 
space (with the radius determined by the effective charge $K_*$)
and show    that  by doing perturbation theory in
the small parameter 
${P^2\ov K_*} \ll 1$  one can match it onto  the KT  asymptotics
at large distances ($u\to 0$ or $\r \to \infty$).
The crucial point will be that $O({P^2\ov K_*})$ 
perturbations will be regular at small distances.
This will be exactly  parallel to the discussion of 
the non-extremal
case in \ghkt\ where the starting point in the IR was 
 a regular non-extremal D3-brane 
 (black hole  in $AdS_5$)  solution  with 
 large (above critical)  Hawking temperature.

 We shall assume, for notational simplicity, 
  that the value  of the radius $L$ 
of  the  short-distance limit space $AdS_5 \times T^{1,1}$  is 1.
That corresponds to the choice of 
the normalizations  where  
the effective 3-brane charge is ($g_s=1$) 
\eqn\pok{ K_* = 4\ , \ \ \ \ \ { \rm i.e. } 
\ \ \ \ \ \ \     L=1 \ . }
In discussing the $O(P^2)$ deformation  it will be useful 
to compare the three possible regular starting points --
the $AdS_5 \times T^{1,1}$ space in the three different
parametrizations, where the constant radial slice 
is $R^4$, \rs\ and \sf respectively:
\eqn\glop{
ds^2_{10}= e^{2\r}( dR^4)^2 + {  d\r^2} + (dT^{1,1})^2 \ ,\ \ \ \  \ \ \ \  \ \ \
-\infty < \r < \infty \ ,  }
\eqn\glos{
ds^2_{10}= \cosh^2 \r \ dX_0^2 + \sinh^2 \r   \ (dS^3)^2  
 + d\r^2  + (dT^{1,1})^2 \ ,  \ \ \ \  \ \ \ \ \ \ \
0< \r < \infty \ ,  }
\eqn\gloss{ 
ds^2_{10}= \sinh^2 \r  \ (dS^4)^2 \ + d\r^2 + (dT^{1,1})^2 \ ,
  \ \ \ \ \ \ \ \ 
\ \ \
0< \r < \infty \ , }
While these three spaces (with   the Euclidean 
$AdS_5$ metric written
in Poincare, global and ``hyperboloid''
parametrizations) 
are related  locally by  the coordinate transformations, 
these involve changing   all five of the coordinates, 
i.e. the radial, but also the angular ones. 
It is the assumption that the 10-d deformation \mett\ 
of the factorized metrics \glop,\glos\ and \gloss\
  under switching on 
 the 3-form flux  \har\  depends 
only on the corresponding 
radial coordinate $\r$ (which is different in the three cases)
that 
makes the resulting solutions different.
Since the 10-d 
metric is no longer a direct product, 
different   choices of the radial coordinate (or of  
 the metric on the $\r=\const$ slice) 
 lead to  inequivalent 10-d equations and thus inequivalent
 D3-brane  solutions no longer related by a local coordinate
 transformation {\it  beyond} 
  the  short-distance  $AdS_5 \times T^{1,1}$
 limit.

In particular, while the Poincare patch  metric \glop\
leads to the KT solution which is  singular in the 
IR (for $\r \to -\infty$), that singularity  is resolved  in the 
$R\times S^3$  and $S^4$  where the $\r\to 0$ limit is described by 
 \glos\ and \gloss,  respectively.

%%%%%%%%%%%%%%%%%%%%%%%%%%%%%%%%%%%%%%%%%%%%%%%%%%%%%%%
\newsec{Action for equations of  radial evolution and special cases }
%%%%%%%%%%%%%%%%%%%%%%%%%%%%%%%%%%%%%%%

As in \refs{\KT,\fiv,\ghkt} 
  we shall first derive the 1-d effective action  that generates the 
  equations 
for the radial evolution of unknown functions.
Computing the scalar curvature  for the metric \mett\ 
we find that  in the \rs\ case \mett,\hih\ 
 $\int d^{10} x \sqrt G  R$
is proportional to  $I_{gr}= \int du \ L_{gr}$, where 
\eqn\gre{
L_{gr}{(R\times S^3)} =
  5 y'^2 - 3 x'^2 - 2 z'^2  - 5
 w'^2  + { 3\ov 2 }  e^{-2x+10y -4z} 
+  e^{8y} ( 6 e^{-2w} - e^{-12 w} )   \ . 
}
The corresponding expression  in the \sf\  case \mett,\tris\  is 
\eqn\gres{
L_{gr}{(S^4)} = 5 y'^2 - 2 z'^2  - 5
 w'^2  + { 3 }  e^{10y -4z} 
+  e^{8y} ( 6 e^{-2w} - e^{-12 w} )  \ . 
}
Note that  it  can 
be formally   obtained from   \gre\  by  setting\foot{The 
coefficient 2 accounts for the ratio of the values of the
Ricci scalars  of $S^3$ and $S^4$.}
\eqn\kkk{ x =\const \ , \ \ \ \ \ e^{-2x } = 2 \ . } 
The new term in $L_{gr}$ \gre\ compared to the (non)extremal 
$R\times R^3$ case  in 
\refs{\KT,\fiv}  is the  first 
 potential term  that 
reflects the
curvature of $R\to S^3$ (or
 $S^4$).\foot{Its scaling  under shifts of
$x,y,z$  follows directly from the structure of
 the metric 
\mett. Shifting $x$ or $z$ to restore explicitly the inverse 
radius parameter 
of   $S^3$ or $S^4$   as its coefficient, one may then recover 
the $R\times R^3$ case in the  limit  when this parameter goes to zero.
As in \fiv,  in the absence of matter terms 
$w=0$  is a consistent fixed point of the
equations of motion, corresponding to  
 $M_5$  in \mmm\ replaced 
by the standard  $T^{1,1}$. Note also that a 
 special solution of the 
equations $R_{mn}=0$  that follow from this gravitational action
is  $R$ times a cone over $S^3 \times T^{1,1}$ or 
 a cone over $S^4\times T^{1,1}$.}

 The matter part $L_m$ of the 
effective type IIB Lagrangian (contributions 
of the dilaton, 3-form
fields and the 5-form  following from \har,\fiff)
is  essentially the same as  in the KT case 
\KT\  and \fiv\ since  
$L_m$ does not
depend on the function $x$ and the structure of $M_4$. As a result, 
$
L=  L_{gr} + L_m =  T  - V $ , where 
\eqn\tki{
  T =  5 y'^2  - 
 3 x'^2 - 2 z'^2  - 5 w'^2 
- { 1 \ov 8} \P'^2
-  { 1 \ov 4}  e^{-\P +  4z -4y - 4 w } {K'^2\ov 4 P^2}  
\ , 
 } \eqn\lagg{ 
 V = 
  - { 3 \ov 2 } e^{-2x+10y -4z}  -
    e^{8y} ( 6 e^{-2w} - e^{-12 w} )
  +   { 1 \ov 4} e^{\P+  4z + 4y + 4 w } P^2 + 
 { 1 \ov 8}  e^{8z} K^2 \ .
}
The   equations  of motion that follow from $L$ 
should be 
supplemented by  the ``zero-energy"  constraint $T+V=0$. 
As in \ghkt, we will use  
the 5-form flux 
function $K(u) = Q + 2 P f(u)$  instead of $f(u)$ in \har.

The new  potential  term $ e^{-2x+10y -4z} $ in \lagg\ associated with the scalar curvature of
the 
4-space, in general, leads to 
breaking of  supersymmetry 
and thus to  a non-trivial 
modification of the extremal KT solution.
In the non-extremal case discussed in \fiv\ this term was absent
and   the 
 equation for $x$ was simply giving  $x=a u$, with
$a$ being  the non-extremality parameter.
In the \rs\ case  with 
 the function $x$ is no longer a  ``modulus'' --
it  cannot be easily decoupled.
In the \sf\ case  the  new potential term in \gres\ provides 
a non-trivial mixing between the $y,z$.

Let us first consider some  special 
solutions of the equations following from this action.

%%%%%%%%%%%%%%%%%%%%%%%%%%%%%%%%%%%%%%%%%%%%%%%%%%%%%%%
\subsec{Flat  4-space case: \ \ extremal KT solution }
%%%%%%%%%%%%%%%%%%%%%%%%%%%%%%%%%%%%%%%

Let us first recall  the solution in the 
$M_4= R\times R^3$ case, 
(corresponding formally to the ``infinite radius''
limit $x=\infty$ of \lagg).
The crucial observation made  in \KT\ is that  
in the absence of the $e^{-2x+10y -4z} $ term 
 the Lagrangian \tki,\lagg\ admits a superpotential,
 i.e.
$L
= g_{ij} (  \p'^i + g^{ik} \del_k W ) ( \p'^j + g^{jl}
\del_l W) 
  - 2 W'.$
As a result, there exists a 
 special BPS  solution 
satisfying     $  \p'^i + g^{ik} \del_k W  = 0$
and thus also the zero-energy constraint.
In 
 the present case  \refs{\KT,\PTs}
\eqn\supp{
W=   { 1 \ov 4} e^{4y} (3 e^{4 w}  +  2 e^{-6w}  )
 - { 1 \ov 8} e^{4z}  K
\ ,  }
and the  corresponding system  of 1-st order equations
is 
\eqn\bps{
x'=0\ ,\ \ \ \ \ 
y' + { 1 \ov 5} e^{4y}(3 e^{4 w} + 2   e^{-6w} )      
=0 \ , \ \ \ \
\ \ \ \ \
w' - { 3 \ov 5} e^{4y} (e^{4w} -  e^{-6 w})       =0 \
,  }
\eqn\zf{\P'=0\ , \ \ \ \ \ \ \ \ 
 K'  + 2 P^2 e^{\P + 4y+4w}  =0 \
, \ \ \ \ \ \ \ \
\ 
z' + { 1 \ov 4} e^{4z} K =0\ . }
Choosing the special solution $w=0$
 we then  find     \KT\ \foot{ $u= { 1 \ov 4 r^4}$ where $r$ is the
 standard  radial coordinate
 in D3-brane solution.}
\eqn\kot{ x=w=\P=0 \ , \ \ \ \ \   e^{-4y} = 4u \ , \ \  \ \ \ \ \
\ 
\ K= K_0  - {P^2\ov 2} \ln u \ ,   } 
\eqn\ktt{ e^{-4z} = h =  
 h_0 +   (K_0 + {P^2\ov 2} ) u  
 -  {P^2\ov 2}  u \ln u \ ,       }
 where $h_0=0$ if we omit the
 standard 
  asymptotically flat region (as we shall assume below).

%%%%%%%%%%%%%%%%%%%%%%%%%%%%%%%%%%%%%%%%%%%%%%%%%%%%%%%
\subsec{ $ K=\const \ (P=0)$ case: \ \ 
  $AdS_5 \times T^{1,1}$  with $M_4= R \times S^3$\ or\   $M_4=S^4$ }
%%%%%%%%%%%%%%%%%%%%%%%%%%%%%%%%%%%%%%%

Setting first fractional 3-brane flux to zero  $P=0$
(i.e. $K=K_*=\const$ and  also $\P=f=0$)  
we get  from \tki,\lagg:
\eqn\addss{
L=  5 y'^2 - 3 x'^2 - 2 z'^2 -5 w'^2  + 
{ 3\ov 2 } e^{-2x+10y -4z} 
+  e^{8y} ( 6 e^{-2w} - e^{-12 w} )   
-{ 1 \ov 8}K^2_* e^{8z}    \ . 
}Here the first term in the potential  
is the contribution of the
curvature of $S^3$, the second is the curvature of the
($w$-deformed)
$T^{1,1}$ space, and the  last one is the negative 
5-d cosmological 
constant originating from the  5-form flux  contribution.
Shifting $z$ and $x$  we may set the D3-brane charge parameter
 $K_*$ to some fixed value, e.g., 
 $K_*=4$ as in  \pok.
 
Since $w=0$ is an  obvious special solution, 
in this case  we get  
 \eqn\aed{
L=  5 y'^2 - 3 x'^2 - 2 z'^2  + 
 {3\ov 2}  e^{-2x+10y -4z}  +  5 e^{8y}  -  2  e^{8z}    \ . 
}
In the standard ``flat'' D3-brane case, 
i.e. in  the absence of the $e^{-2x+10y -4z}$ term, 
 this system is easily integrated 
giving us 
extremal ($x=0$) or non-extremal ($x=au$) D3-brane 
on conifold solution.
  The case of 
   \eqn\uyo{ y=z \ , } 
 then  corresponds to the $AdS_5 \times T^{1,1} $ 
limit  \glop\
where  the $M_5$ part of the metric \mett\ factorizes.

 In general, while  it is not clear  how to solve 
the system that follows from 
\aed\ analytically, it is easy to see that 
  the 5+5 factorized case  \uyo\
  is still a special 
solution. Here  we end up with
 \eqn\aetd{
L=  3 ( y'^2 -  x'^2   + 
  {1 \ov 2}  e^{-2x+ 6y}  +   e^{8y} )      \ . 
}
The corresponding  equations  have the following solution
\foot{Note that while for $q\not=0$  or $y\not=z$ 
\aed\ does not admit a superpotential, 
  it exists for \aetd\ (cf. \supp) 
$
W = {3\ov 4}( {1 \ov 2}  e^{-2x + 2y} +  e^{4y}).$}
\eqn\sop{
e^{4x} = \tanh \r \ , \ \ \ \ \ \ \ 
e^{4y} = \sinh^3 \r\ \cosh \r \ , \ \ \ \ \ \ \ \ \ 
d\r=- e^{4y} du \ , }
where we have set the  only integration constant 
(the origin of $\r$) to zero.\foot{Here 
$u= \ln \tanh\r + \sinh^{-2} \r$, so that 
$u(\r\to \infty) \to 2 e^{-2\r}$ 
 and 
$u(\to 0) \to { 1 \ov 2 \r^2}$.}  
The metric is thus given by \glos, 
i.e. is  the product of $AdS_5$ in the 
global parametrization   and  $T^{1,1}$ 
(both with scale  $L=1$).
Large $\r$ corresponds to the boundary, and 
small $\r$ -- to the origin of $AdS_5$.

%%%%%%%%%%%%%%%%%%%%%%%%%%%%%%%%%%%%%%%%%%%%%%%%%%%%%%%
%\subsec{\sf case with $P=0$: \ \ 
 % $AdS_5 \times T^{1,1}$   }
%%%%%%%%%%%%%%%%%%%%%%%%%%%%%%%%%%%%%%%
In the   \sf case \gres\ setting  $K=K_*=\const$
gives  (e.g., using \kkk\   in \addss)
\eqn\dss{
L=  5 y'^2 - 2 z'^2 -5 w'^2  + 
{ 3 }  e^{10y -4z} 
+  e^{8y} ( 6 e^{-2w} - e^{-12 w} )   
-   { 1 \ov 8} K^2_*   e^{8z}    \ , 
}
or, for  $w=0$,  and $K_*=4$, 
 \eqn\saed{
L=  5 y'^2  - 2 z'^2  + 
 3  e^{10y -4z}  +  5 e^{8y}  -  2  e^{8z}    \ 
}
The meaning of the three terms in the potential is again the 
curvature of $S^4$, the curvature of $T^{1,1}$ and 
negative cosmological term produced by the $F_5$ flux. 
Equivalently, 
 \eqn\ssees{
L=  3 n'^2 - 30 m'^2  + { 3 } e^{6n } 
+  e^{8n} ( 5 e^{-16m}  - 2  e^{-40m}  ) 
  \ , \ \ \ \     z= n- 5 m \ , \ \ \ \ \    y= n - 2 m \ .
  } 
  In general, this system  does not
   admit  a  superpotential
  (wrapping
 the  Euclidean 3-brane world-volume over $S^4$  breaks
  supersymmetry).
The  special easily solvable  case is  the fixed point $ m=0$, 
i.e. $y=z$ \uyo\ or  
 the case of factorization $M_{10} \to M_5 \times T^{1,1}$. 
Here  we are left  with  just with one function  $y$
satisfying 
the zero-energy constraint (there is thus an obvious
superpotential, cf. \aetd)
\eqn\coon{
y'^2 =    e^{ 6y}  +   e^{8y}   \ , }
so that  
\eqn\mom{
z=y  = \ln   \sinh \r \  , \ \ \ \ \ \ \ \ \ \ \ \ 
  d\r= - e^{4y} du  
    \ , }
  where we again set $\r_0=0$.\foot{Here 
   $ u= { \cosh \r \ ( 1- 2 
  \sinh^2 \r) \ov 3 \sinh^3 \r
  } + {2 \ov 3} $, \  so that   $ u(\r\to 0) \to { 1 \ov 3 \r^3}, \ 
 \  u(\r\to \infty) \to  4 e^{-4\r} . $ } 
  Then 
  the metric becomes equal to \gloss, with the $AdS_5$ 
  part written  in the parametrization where 
the topology of the radial slices is $S^4$.

It is useful to  stress again that the three 
$AdS_5 \times T^{1,1}$ 
metrics \glop,\glos, 
and \gloss, though related locally by the  coordinate transformations, 
are obtained from {\it inequivalent} 
1-d actions. This reflects inequivalence of the corresponding 
radial coordinates, and  leads also to  
 very different properties of the 
corresponding fractional brane ($P\not=0$) 
deformations of these 
backgrounds discussed below.

%%%%%%%%%%%%%%%%%%%%%%%%%%%%%%%%%%%%%%%%%
\newsec{Strategy of finding $P\not=0$  solution  and $S^4$ case }
%%%%%%%%%%%%%%%%%%%%%%%%%%%%%%%%%%%%%%%%%%%%%%%%%

Being unable to solve the system of equations that follows 
from \tki,\lagg\  in general, 
we need to resort to perturbation theory 
similar
to the one used in \ghkt.
Our aim will be to show that 
starting from the asymptotic  KT geometry  at large $\r$ 
 one may 
smoothly interpolate  to a 
{\it regular}
 $AdS_5 \times T^{1,1}$  geometry  (with large enough 
effective charge $K_* \gg P^2$)
  at small $\r$ 
  with the metric  having a non-trivial 
scalar curvature  of $\r=\const$ slices, i.e.
\glos\ or \gloss.

Following the same strategy as used in in \ghkt\
in finite temperature case 
we shall start  with the $AdS_5 \times T^{1,1}$ background \gloss\ 
expected to be a good approximation in the small 
$\r$ region if $K_*= K(\r\to 0)$ is sufficiently large, 
and solve the  supergravity equations perturbatively 
to leading order in ${P^2\ov K_*}\ll 1 $.
We shall see that the leading deformation of the 
$AdS_5 \times T^{1,1}$
background will be {\it regular} at small $\r$.

If one starts  instead 
 with  the ``flat''   $AdS_5 \times T^{1,1}$
  metric \glop, such perturbative expansion 
reproduces  the exact form of the  KT solution 
 already at the first order of perturbation theory in $P^2\ov K_*$
 (note that the correction terms  in \kot,\ktt\ are linear in $P^2$).
 Here,  however,  the perturbation
 (and the exact solution) is  {\it singular} 
  in the short-distance region (which 
  in the case of \glop\ corresponds  to $\r\to -\infty$).
As was explained in \ghkt, introducing non-extremality 
(i.e. replacing  $AdS_5$ by the black hole background 
with  sufficiently high  temperature) resolves 
 the singularity, making 
the perturbative solution regular. 
  We shall see that a similar resolution is 
  provided by the curvature of the ``parallel'' 3-brane directions.

As was already mentioned above, 
to simplify the presentation 
 we shall assume that the value of  the 5-form flux at $\r\to 0$ 
is fixed  as in \pok,  so that the radius of  $AdS_5$ is 1
as in \glop--\gloss.
The  expansion  parameter is  then simply $P^2$.

The  full  
system of 2-nd order equations 
following from \tki,\lagg\ in 
the   \rs\  case is similar  to the one in \ghkt\
 \eqn\xexx{
  x'' - {1\ov 2}  e^{-2x-4z + 10y} =0 \ , }
\eqn\yyy{ 10y'' - 8 e^{8y} (6 e^{-2w} - e^{-12 w})  - 30 x''
 + \P''
=0 \ , 
}
\eqn\yuw{
10w'' - 12 e^{8y} ( e^{-2w} - e^{-12 w})   - \P''
=0 \ , } 
\eqn\ppp{
\P''    + e^{-\P + 4z - 4y-4w} ( {K'^2\ov 4 P^2}-  e^{2 \P + 8
y+8w} P^2)=0 \
,
}
\eqn\zzy{
4z'' -  K^2  e^{8z}
 - e^{-\P + 4z - 4y-4w} ( {K'^2\ov 4 P^2} +  e^{2 \P + 8 y+8w} P^2)
  - 12 x''  =0\ , 
}
\eqn\fef{
(e^{-\P + 4z - 4y-4w} K')' - 2P^2 K e^{8z} =0 \ . 
}
The integration constants are 
subject  to the  zero-energy constraint $T+V=0$.
It is easy to see that because of the extra $S^3$-curvature term 
$e^{-2x-4z + 10y}$  in the potential 
 this  system does {not} (in contrast to the non-extremal
 case \Buch)
admit a special solution with constant dilaton 
and self-dual 3-forms.\foot{If we set 
$K'^2 -  4P^3 e^{2 \P + 8 y+8w}=0$, 
i.e.~$K'= -  2P^2  e^{ \P + 4 y+4w} $, then \fef\ 
implies that $z$ should be subject  to the 
first-order equation in
\zf, but this is not consistent with \zzy\ unless $x''=0$.}
In \fiv\
we needed  to relax this 1-st order system 
to get a non-singular non-extremal solution.
Here we do not  have a choice --  all functions
(in particular, $w$)  are   to be non-trivial
in general.

In the \sf\ case we get instead\foot{This system is related 
to the \rs one  by 
setting  $e^{-2x}=2$ in \xexx--\fef\ 
 after using \xexx\ in  \yyy, \zzy.}
\eqn\yy{ 10y'' - 8 e^{8y} (6 e^{-2w} - e^{-12 w})  -
 30    e^{10y-4z }
 + \P''
=0 \ , 
}
\eqn\uw{
10w'' - 12 e^{8y} ( e^{-2w} - e^{-12 w})   - \P''
=0 \ , } 
\eqn\pp{
\P''    + e^{-\P + 4z - 4y-4w} ({K'^2\ov 4 P^2} - 
 e^{2 \P + 8 y+8w} P^2)=0 \ ,
}
\eqn\zy{
4z'' -  K^2  e^{8z}
 - e^{-\P + 4z - 4y-4w} ( {K'^2\ov 4 P^2} +  
 e^{2 \P + 8 y+8w} P^2)
  - 12    e^{10y-4z }   =0\ , 
}
\eqn\ff{
(e^{-\P + 4z - 4y-4w} K')' - 2P^2 K e^{8z} =0 \ , 
}
with the zero-energy constraint 
$$5  y'^2    - 2 z'^2  - 5 w'^2 - { 1 \ov 8} \P'^2 
- { 1 \ov 4}  e^{-\P +  4z -4y - 4 w }{K'^2\ov 4 P^2}  $$ 
\eqn\coln{  -  \ { 3 }   e^{10y -4z}
 -    e^{8y} ( 6 e^{-2w} - e^{-12 w} )
 +  { 1 \ov 4} e^{\P+  4z + 4y + 4 w } P^2 + 
 { 1 \ov 8}  e^{8z} K^2   = 0  \ . 
}
This system is   simpler than in  the \rs\ case, 
and in  the remainder of this section 
 we shall concentrate on its
solution for the first $O(P^2)$ deformation away from 
the $AdS_5 \times T^{1,1}$ metric \gloss.

%%%%%%%%%%%%%%%%%%%%%%%%%%%
\subsec{Asymptotics of regular $S^4$  solution}
%%%%%%%%%%%%%%%%%%%%%%%%%%%%%%%%%%%%%%%%%%%%%%%%%%%%%%%%%

Let us first discuss the expected 
behavior of the solution in
the two asymptotic regions: $\r \to 0$ ($u\rightarrow \infty $)
 and
$\r \to \infty$ ($u\rightarrow 0$), 
i.e. in the short-distance 
 and long-distance limits in 10-d space.
We would like  the solution to have a regular short-distance 
limit  which has the form \gloss\
 (up to possible constant rescalings) 
\eqn\boun{ 
\r\to 0: \ \ \  \ \   y\to  \ln \r  + y_*  \  , \ \  
\ \  \  
z\to   \ln \r  + z_*
, \ \ \  \ \    w\to w_*\ , \ \  \ \   
\P\to \P_* \ ,  \ \  \ \ \ \ K\to  K_*\ . }
At large distances 
($\r\to \infty $) the  solution is expected  to  approach 
the extremal
KT solution \kot,\ktt, i.e. (note that according to 
\mom\  $u(\r\to \infty) \to  4 e^{-4 \r} $)
\eqn\lla{
\r\to \infty: \ \ \ \ \ 
w\to 0 , \ \ \ \P\ \to\  0  , \ \ \  
e^y \to  \ha e^\r    , \ \ \ 
K \to 2P^2 \r \ ,
 \ \ \  e^{-4z} \to  {8P^2 } \r e^{-4 \r} 
 \ .
}
To  demonstrate 
the existence of a regular solution  which interpolates between 
these two asymptotics 
we shall  start with 
 \gloss\ 
which is valid for
$P=0$, and find its deformation order by order   in $P^2$.
We shall see that  (under a proper choice of integration constants)
 the leading    $O(P^2)$       perturbations
 are {\it regular}  
at $\r\to 0$, so that we  indeed 
 match onto 
the short-distance  asymptotics \boun.
It turns out 
that the leading $O(P^2)$ correction 
 is already enough 
to match onto the expected 
KT long-distance asymptotics \lla. 
 
Our ansatz for  the leading  perturbative 
 solution that differs from  
 \gloss\ by the $O(P^2)$ terms will be 
\eqn\seeq{
K = 4 + 2 P^2 k(\r)
 \ , \ \ \ \ \ \P  = P^2 \p(\r) \ , \ \  \ \ 
 w = P^2 \om (\r) \ , }
 \eqn\joi{
\ y= y_0(\r)   +  P^2 \xi (\r) \ ,\ \ \ \ \ \ \
 \ \ \   z=  y_0(\r)   +  P^2
\eta (\r) \ , \ \ \ \   \  \ \ \ \ \ 
  y_0(\r) \equiv \ln \sinh \r \ , 
}
We shall look for  solutions  for the 
 perturbations  $ k, \p, \om, \xi, \eta$ 
   which are regular at 
$\r\to 0$
\eqn\zerr{
\r\to 0: \ \ \ \ \  \ \ \ \ 
  k, \p, \om, \xi, \eta \ \ \to \ \ \const \ ,  }
in agreement with \boun. 
We will find then  that at  large $ \r$  the solution matches  onto 
the KT asymptotics \lla\ 
\eqn\matching{
\r\to \infty: \ \ \ \ \ 
\om,  \p,  \xi \to  0\ , \  \quad \  \ 
k  \to \r  \ , \ \ \ \ \ \eta \to - {1\over 8}\r\ 
 \ . }

%%%%%%%%%%%%%%%%%%%%%%%%%%%%%%%%%%
\subsec{Solution for  $O(P^2)$ perturbations}
%%%%%%%%%%%%%%%%%%%%%%%%%%%%%%%%%%

Substituting \seeq\ into the  system \yy--\coln\
we get 
\eqn\yyu{ 10\xi'' - 320  e^{8y_0} \xi 
    - 60 e^{6y_0} (5\xi - 2 \eta) +  \p''  +O(P^2)    =0 \ , }
\eqn\typs{ 
10\om''  -  120 e^{8y_0} \om - \p'' +   O(P^2)  =0
\ , }
\eqn\pu{
\p'' + k'^2 - e^{  8 y_0}+ O(P^2)=0 \ , }
\eqn\zy{
4\eta'' - 128 e^{8y_0}\eta -24 e^{6y_0} (5 \xi - 2 \eta) 
 - ( 16k +1) e^{8y_0}  -    k'^2 
   +O(P^2) =0\ , 
}
\eqn\fefh{
k'' -  4 e^{8y_0}  + O(P^2)=0 \  , 
}
\eqn\cvo{
 2y_0' (5 \xi'  - 2  \eta' ) 
- { 1 \ov 4}   k'^2  + e^{8 y_0} [  { 1 \ov 4}   + 2  k  
- 8 (5 \xi  - 2    \eta ) ]  - 6 e^{6y_0} (5 \xi  - 2    \eta )
+ O(P^2) = 0 \ . }
Here prime stands for the derivative over $u$, 
with $du= - e^{-4 y_0} d \r$ (see \mom).
Changing to  the derivatives over $\r$ we 
finish with 
\eqn\fefm{
k'' + 4 y_0' k'  -  4 =0 \  ,  \ \ \ \ \ \ \ 
 \ k' \equiv  { d k\ov d\r} = - 
 e^{-4y_0} { dk \ov du} \ , \ \  y_0'= \coth \r\ , 
}
\eqn\pun{
\p'' + 4 y_0' \p'  + k'^2 - 1 =0 \ , }
\eqn\yps{ 
\om'' + 4 y_0' \om' -  12 \om  +   {1\ov 10} ( k'^2 - 1 )  =0 
\ , }
\eqn\yyu{ \xi'' +    4 y_0' \xi'
- 32  \xi   - 6 e^{-2y_0} (5 \xi - 2 \eta)   -   {1\ov 10}(
k'^2 -  1 )   =0 \ , }
\eqn\zy{
\eta'' +   4 y_0' \eta'  - 32   \eta 
 -  6 e^{-2y_0} (5 \xi - 2 \eta)  
   - {1 \ov 4} (  k'^2 +1 + 16 k )  =0\ , 
}
\eqn\cvo{
 y_0' (5\xi'  - 2  \eta' ) - (3 e^{-2 y_0} + 4)(5\xi  -2   \eta )
- { 1 \ov 8}  ( k'^2  -1 - 8k )    = 0 \ . }
The deformation of the background 
is thus driven by the  perturbation $k(\r)$ of the
effective 3-brane charge $K$; solving for $k(\r)$ first we then 
determine the source terms in the 
linear equations for the remaining perturbations. 
The equation \fefm\ is readily solved:
\eqn\jop{
k= - {5 \ov 6}  + \r \coth \r \ ( 1 - { 1 \ov 2 \sinh^2\r})  + 
{ 1 \ov 2 \sinh^2\r}\ , } 
where we have fixed the two 
integration constants so that to satisfy 
the condition \zerr\ of regularity at small $\r$: \  $k(0)=0$.
Indeed, $k (\r\to 0) \to { 2 \ov 5} \r^2  + O(\r^4)$.
We   also get the expected 
matching onto the KT asymptotics \matching:
 $k(\r\to \infty) \to \r$.

The solution  for the dilaton  perturbation  \pun\ is then:
\eqn\pu{
\p= { 13 \ov 72} -  { 1 \ov 2 \sinh^2 \r}
+   { 3\r^2   +2 \r \coth \r \  -1 \ov 8 \sinh^4 \r}
-  { \r^2  \ov 8 \sinh^6 \r} \ , 
}
where again we have  fixed  the integration constants so that 
   to have the  regularity at small $\r$, 
   \ $\p(\r \to 0) ={ \r^2\ov 10} + O(\r^4)$.
At large $\r$ the dilaton perturbation 
exponentially approaches zero, in agreement with \matching.

The three  equations    for the gravitational 
perturbations  $\om,\xi,\eta$  have a similar  structure
(as was also the case in \ghkt).
For $\om$ we  get 
\eqn\psu{
\om'' + 4\coth\r\  \om' -  12  \om   + q (\r)  =0  \ ,}
$$  q \equiv {1\over 10} ( k'^2  -1 ) ={1\over
10}\bigg[
{ (12 \r  - 8 \sinh 2 \r + \sinh 4\r )^2  \ov 640\sinh ^8\r }   - 1\bigg]
\ . $$ 
Note that the source term  is regular at 
small $\r$: $q(\r\to 0)  \to -{ 1 \ov 10}  + { 8 \ov 125} \r^2 + ...$, 
and $q(\r\to \infty)  \to  { 12 \ov 5} e^{-2\r}  + O(\r e^{-4\r})$.
As a result,  this equation has a regular solution 
near $\r=0$:\  $\w= w_* + 
({6\ov 5} w_* + { 1 \ov 100})   \r^2 + ...$.\foot{ 
Note   that \psu\ can be put also in the following form
(which is of the same type that appeared in \ghkt)\ 
$
\td \om''  -  2 (6 + { \cosh 2 \r\ov \sinh^2 \r}) 
  \td \om   + \sinh^2\r\ q (\r)  =0  \ ,
\ \ \ \ \ \   
 \om = e^{-2y_0} \td \om = \sinh^{-2} \r\ \td \om .$  }
 It is easy to see (following the analysis in \ghkt\ or by numerical
 integration) that this regular short-distance 
 asymptotics  is smoothly connected
 to the  long-distance asymptotics $w \to { 3 \ov 20} 
 e^{-2\r}\to 0$.

The equations for $\xi$ \yyu\ and $\eta$ \zy\ are coupled
though the $5\xi-2 \eta$ term, 
but the equation for this combination can be easily integrated.
In fact, its solution  can be  found from the constraint
\cvo:
\eqn\cvot{
  \nu'   +\   p_1 (\r) \ \nu\  +\  p_2(\r) =0 \ , \ \ \ \ \ \ \ \ \ \ \ 
  \nu\equiv  5 \xi -2 \eta \ , }
  $$
p_1\equiv   - ({ 3\ov \sinh^2\r }  + 4)\tanh \r\ , \ \ \ \ \  \ \ \ 
p_2 \equiv  -{ 1 \ov 8} \tanh\r  \ ( k'^2  -1 -8k) \ .  $$
This gives:
\eqn\hoj{
\nu = - S(\r)   \int d\r \  S^{-1}(\r) \    p_2(\r)    \ , 
\ \ \ \ \ \   
 S\equiv  e^{- \int d\r\   p_1 (\r)}  =  \sinh^3 \r \ \cosh \r \ . 
}
The resulting expression for $\nu$ (given in terms of
the  dilogarithm  function)  is  regular at small $\r$: 
\ $ \nu (\r\to 0) =  {1\ov 8}  \r^2+ 
 + O(\r^3)$,  
 while for large $\r$ we get $\nu \to {1 \ov 4} \r$, 
 in agreement with  \matching.
 
We are left with only one equation to solve --
for $\xi$ (or for $\eta$)  
\psu\  
\eqn\yeu{ \xi'' +    4 \coth \r \  \xi' - 32  \xi   + v(\r)  =0\ , }
  $$ 
v\equiv   -\tanh\r\ [ 
{ 6 \ov \sinh^2\r}  \nu    +   {1\ov 10}(k'^2 -  1 )]    \ . $$
Its analysis is the same as for \psu.
Since the source $v$  here is again 
regular  at $\r\to 0$: 
$v= v_0 +  O(\r^2), \ v_0= -{13\ov 20}$, 
the solution for $\xi$ is also regular, 
$\xi= \xi_*  + ( { 16 \ov 5}  \xi_*  - {v_0\ov 10} ) \r^2  + O(\r^4)$. 
As in the case of \psu,  we  are also  
 able to  connect this  to the required large $\r$ asymptotics 
  \matching, i.e. $\xi \sim e^{-2\r}\to 0$. 

We conclude that  both the matter 
the  gravitational 
perturbations are {\it regular } at  small $\r$, 
and match onto the KT solution at large $\r$.

It is instructive to see explicitly why  replacing 
$S^4$  by $R^4$, i.e. going back to the original KT ansatz, 
gives singular solution, i.e. why 
repeating the above perturbative 
analysis in the $R^4$ case  leads to  singular $O(P^2)$ 
corrections, even though the starting point --  $AdS_5 \times T^{1,1}$
space 
in Poincare coordinates \glop\ is non-singular (see also \ghkt). 
 Omitting the potential term associated  with the curvature of  $S^4$ 
in \yy,\zy,\coln\  and using the ansatz \seeq,\joi\ with 
$y_0= - {1 \ov 4} \ln (4u)  = \r$ (cf. \kot,\glop)
we finish with the following system of equations 
that replaces \fefm--\cvo\ ($y_0'=1$) \foot{The derivative here is 
over $\r$  that here takes values $-\infty < \r < \infty$, 
with $\r \to -\infty$ being the short-distance limit.}
\eqn\efm{
k'' + 4  k'  -  4 =0 \  ,  \ \ \ \ \ \ \  
\p'' + 4 \p'  + k'^2 - 1 =0 \ , }
\eqn\ypos{ 
\om'' + 4  \om' -  12 \om  +   {1\ov 10} ( k'^2 - 1 )  =0 
\ ,  \ \ \ \ \  \xi'' +    4  \xi'
- 32  \xi   -  -   {1\ov 10}(k'^2 -  1 )   =0 \ , }
\eqn\zyo{
\eta'' +   4  \eta'  - 32   \eta  
   - {1 \ov 4} (  k'^2 +1 + 16 k )  =0\ , 
}
\eqn\cvoo{
5\xi'  - 2  \eta'  - 4 (5\xi  -2   \eta )
- { 1 \ov 8}  ( k'^2  -1 - 8k )    = 0 \ . }
Fixing the integration constants so that to achieve 
maximal  possible regularity of  functions at 
 at $\r=-\infty$ we get 
\eqn\reso{
k =\r \ , \ \   \p=0 \ ,  \  \ \om=0 \ , \ \ \ \ 
\eta = - { 1 \ov 32} - { 1 \ov 8} \r \ , \ \ \ 
\xi=0 \ .  }
This reproduces \kot,\ktt\ (note that $e^{-4z} =
 e^{-4y_0}(1 + P^2 \eta + ...)$), and thus leads to a
 short-distance singularity at $\r\to -\infty$.
 It is  the  singular behaviour of the ``source function'' 
 $k$  that translates into  the singularity of the gravitational 
 perturbation $\eta$. At the same time, in the 
 non-extremal case  in \ghkt\ and in the 
 present  $S^4$ case  \jop\ (and \rs case discussed below) 
 the function $k$   has {\it regular} short-distance limit.

%%%%%%%%%%%%%%%%%%%%%%%%%%%
\newsec{$P\not=0$  solution in  $R\times S^3$ case} 
%%%%%%%%%%%%%%%%%%%%%%%%%%%%%%%%%%%%%%%%%%%%%%%%%%%%%%%%%

The case of compactification 
on $S^3$ though technically more 
complicated, can be analyzed analogously to the $S^4$ case.
We will construct a smooth supergravity RG flow interpolating between 
a conformal compactification of the KW geometry at the origin,
and the asymptotically KT geometry to the leading order in $P^2$. 
The full second order system is given by\foot{The integration 
constants are subject to the zero-energy constraint as explained 
above.} \xexx\ - \fef. The starting point for the deformation 
by the 3-form fluxes is the $AdS_5\times T^{1,1}$ space in
the  global 
parametrization \glos. In what follows we will use 
the radial coordinate $t$ related to $\rho$ in \glos\ as 
\eqn\rhot{
t=\tanh^2\rho\ , 
}
and to $u$ in \xexx--\fef\ as 
\eqn\ut{
{du\over dt}={e^{z-5 y}\over 2\sqrt{t} (1-t)} \ .  
}
Here  $t\to 0_+$ and  $t\to 1_-$ are the short-distance and the long 
distance limits of the 10-d space, respectively.\foot{These are 
correspondingly the IR and the UV regimes of the holographically dual 
gauge theory.}
Let us also  introduce the functions 
\eqn\redef{
f_1=e^{12 x -4 z}, \qquad f_2=t^2 e^{-4 z -4x} 
 , \qquad
f_3=e^{4y-16w-4 z },\qquad f_4=e^{4y+4w-4 z}\ , 
}
so that the deformed 10-d metric \mett\  takes the form 
 $$
ds_{10E}^2=f_1^{-1/2} dX_0^2+t\
f_2^{-1/2}\left(dS^3\right)^2+{dt^2\over 4 
t(1-t)^2}
$$\eqn\newmett{
+\ f_3^{1/2} e_{\psi}^2+f_4^{1/2} \left(e_{\theta_1}^2+e_{\phi_1}^2+
e_{\theta_2}^2+e_{\phi_2}^2\right) \ . 
}
The reason for the   redefinitions \ut , \redef\ is that 
using $f_i(t)$ it  is  easier 
to construct perturbative in $P^2$ solution to  \xexx\ - \fef .
For $P=0$ we recover the  $AdS_5\times T^{1,1}$ space in the global
parametrization \glos\ 
\eqn\pzero{
f_1=f_2=(1-t)^2,\qquad f_3=f_4=1,
}  
with unit radius corresponding to the choice of $K=4$.

Our anzatz for a perturbative solution that differs from
\pzero\ 
by $O(P^2)$ terms will be similar to \seeq,\joi\ 
\eqn\ppzero{
f_1(t)=(1-t)^2+P^2 \F_1(t),\qquad f_2(t)=(1-t)^2+P^2 \F_2(t)\ , 
}
$$
f_3(t)=1+P^2 \F_3(t),\qquad f_4(t)=1+P^2 \F_4(t)\ , 
$$
$$
K(t)=4+2 P^2 k(t),\qquad \Phi(t)=P^2\phi(t)\ . 
$$
{}From \newmett\ it is clear that to avoid a  singularity 
in the metric at 
$t\to 0_+$ we should have
\eqn\boundzero{
\F_2(t)\to 0,\qquad \ \ \ \   \F_{1,3,4}(t)\to {\rm const}\ . 
}
Also, to reproduce the  $P=0$  values of the dilaton 
and of the regular D3-brane charge $K$ at $t=0$ we shall assume  
that 
\eqn\bondzt{
\phi(t)\to 0\ ,\ \ \ \ \qquad k(t)\to 0\  .
}
At large distances ($t\to 1_-$) the solution is 
expected to approach the extremal KT solution \kot, \ktt\ 
\eqn\bonone{
\F_1(t)\to  2 k(t) e^{-4 k(t)}\ ,\qquad \ \ \ \ \
\F_2(t)\to  2 k(t) e^{-4 k(t)} \ , 
} 
$$
\F_3(t)\to  {1\over 2} k(t),\qquad  \F_4(t)\to  {1\over 2} k(t)
\ , \ \ \ \  
\phi(t)\to 0,\qquad k(t)\to +\infty\ . 
$$
Notice that because $k(t\to 1_-)\to +\infty $, the perturbative 
expansion \ppzero\ necessarily breaks down  there, so 
that,  strictly 
speaking,  we should not expect 
to reproduce the precise form of the 
KT asymptotics \bonone.
This is indeed what we will find. 
We will  recover asymptotically the warped product of 
the two factors $R\times S^3$
(with a finite $S^3$) and $T^{1,1}$, with the 
warp factors differing from the corresponding ones 
in the KT geometry 
by subleading logarithmic corrections. 
The same phenomenon was 
also observed in \ghkt.  
 
Now, changing the radial coordinate 
according to \ut , performing the 
redefinitions \redef\ in \xexx\ - \fef , and substituting the 
expansion \ppzero\ into the resulting system of equations, 
we obtain a coupled system of second-order  equations
 for $\F_{1,2,3,4}(t),
\phi(t), k(t)$. In particular, for $k(t)$ we find
\eqn\kk{
t (1-t)^2 k''+(1-t)(2-t) k'-1=0\ . 
}  
The solution of \kk\ with  the correct  boundary conditions is 
\eqn\kks{
k(t)=-{1\over 2} \ln(1-t)\ . 
} 
For the dilaton perturbation we find
\eqn\dilsol{
t(t-1)^2 \p''+(1-t)(2-t)\p'+ {1\over 4}(t-1) =0\ , 
} 
and its appropriate solution  is 
\eqn\Ds{
\phi(t)=-{1\over 4 t}\left[t\ {\rm Li}_2(t)
+ \ln(1-t) ( 1 - t +\ln t  ) \right]\ . 
}
Next, let us consider the  equations for the $\F_3$ and $\F_4$. 
Introducing 
\eqn\fdef{
\varphi_{34}(t) \equiv   \F_4- \F_3 \ , 
}
 we obtain (using the already determined functions)
\eqn\feq{
2 t(1-t)^2 \varphi_{34}''+2(1-t)(2-t) \varphi_{34}' -{2\over 3} \varphi_{34}
+(t-1)=0 \ . 
}
The solution of \feq\ with the correct asymptotics is 
\eqn\fsol{
\varphi_{34}(t)=
{t+ 2\ov 2 (1-t) }  [ {\rm Li}_2(t) + \ln(1-t)\ln t] 
-{5t+1 \over 4 t} \ln(1-t)- { 3 \ov 2}  \ . } 
Substituting the already determined functions into the equation 
for $\F_3$ we find 
$$(1-t)^4 \big({ t^2\over 1-t}\F_3'\big)'-8 t (1-t)
\F_3 + {1\over
4}t (t^2 - 28 t + 27)  
$$
\eqn\ffoure{ 
-\ 2  t(t+2)  \ln(1-t)\ln t   + 2(7 t^2-6t -1)  \ln(1-t)  -4
 (t^2+2) {\rm Li}_2\
t =0  . } 
Though \ffoure\ looks complicated, the  general solution can still 
be found 
\eqn\ffourgen{
\F_3(t)={1\over 12 t (1-t)^2}\left[I_1(t)+I_2(t)\right]+
 {t^2+6 t +3 \over (1-t)^2} (\a_1 + 3 \a_2 \ln t )  
+\a_2 {51 t^2 +48 t +1 \over 
t (1-t)^2}\ , } 
where 
$$
I_1(t)=-t (t^2+6 t+3)\int_0^t {dx\ov x(1-x)^6}
\! \bigl[51\,{x}^{2}+48\,x+1+3x (x^2 + 6x + 3 ) \ln x \bigr]
$$\eqn\ione{
\times \big[x^3- 3 x^2 + 27x        +4 
 (7x^2 - 6x-1 )  \ln(1-x)-8 x(x+2)
 \ln(x)\ln(1-x) 
-24 x^2 \,{\rm Li}_2(x)    \big],   } 
and 
\eqn\itwo{
I_2(t)=[51 t^2+48 t +1 +3t ( t^2 + 6t + 3) \ln t ]\int_0^{t} {dx\ov
(1-x)^6}
\!
\big(x^2+6 x+3\big) 
}
$$
\times \big[x^3-28 x^2+ 27 x + 4 (7x^2 + 6x - 1) 
 \ln(1-x)-8x (x+2) \ln(x) \ln(1-x) - 24 x^2  {\rm Li}_2(x)       \big]\ . 
$$
Both integration constants $\a_1$ and $\a_2$ are uniquely 
fixed by the boundary conditions. For  $t\to 0$ we find
\eqn\ffourzero{
\F_3(t)={\a_2\over t}+\a_2\left(50+9\ln t\right)+3\a_1
+\a_2\ O\left(t\ln t\right)+O\left(t\right)\ . 
} 
{}From \ffourzero\ we see that the analyticity of $\F_3$ at the origin
requires  $ \a_2=0$. 
In the limit  $t\to 1_-$ we  get 
\eqn\ffourone{
\F_3(t)={10s\over (1-t)^2}-{8s\over (1-t)}+\ s+{1\over 8}
- {1\over 4}\ln(1-t)\ 
+O\left((1-t)\ln(1-t)\right) \ , 
}
where 
\eqn\etad{
s \equiv\a_1+{1\over 120} [ I_1(t\to 1_-)+I_2(t\to 1_-)] \ . 
}
It is straightforward to verify that the sum $I_1(t\to 1_-)+I_2(t\to
1_-)$ 
is actually finite.\foot{Numerically,  we find that 
$I_1(t\to 1_-)+I_2(t\to
1_-)\approx -7.753297.$}
{}From \ffourone\ we conclude that to get the KT asymptotic for the 
$\F_3$ as given by \bonone\ we have to tune $s=0$.
Then  \etad\  uniquely fixes the coefficient $\a_1$. 

We did not find the 
exact analytical solutions for $\F_1,\F_2$, but it is possible to show
that  the 
 regularity at $t\to 0$ fixes all the integration constants but one.
In general, one  finds
\eqn\fffone{
\F_1(t)=\gamma  (1-t)^2+\sum_{i=1}^\infty d_{1i} t^i\ , \ \ \ \ \ \ \ 
\F_2(t)=\sum_{i=1}^\infty d_{2i} t^i\ , 
}
where $d_{1i}$ and $d_{3i}$  are some (uniquely) determined 
coefficients 
and $\g$ is an arbitrary integration constant. The 
 presence of
 $\g $ reflects  the freedom of rescaling of the 
time   coordinate $X_0$ in \newmett.
This arbitrary constant has no effect on the UV ($t\to 1_-$) 
asymptotic, where we find 
\eqn\fasym{
\F_1(t)\to {1\over 16} (1-t)^2 \left[\ln (1-t)\right]^2\ , 
\ \ \ \ \ \ \ \ 
\F_2(t)\to  \F_1(t) \ . 
}
Unlike the solution for the $\F_3$-perturbation,\foot{Recall
 that this function
determines the  asymptotic  warp factor of the $T^{1,1}$ 
space 
in the $R\times S^3$ 
compactification of the KT geometry.} which precisely 
reproduces the corresponding KT asymptotic, the precise form 
of the KT asymptotics 
for $\F_1,\ \F_2$ would be  (see  \bonone ,  \kks)
\eqn\fasymex{
\F_1(t)\to - (1-t)^2 \ln (1-t)\ , 
\ \ \ \ \ \ \ \ 
\F_2(t)\to  \F_1(t) \ . 
}
The (subleading) difference between \fasym\ and \fasymex\ should not be 
surprising. Much like what happens 
in the nonextremal deformation of the KT solution 
\ghkt, our perturbative expansion breaks down at  $t\to 1_-$.

%%%%%%%%%%%%%%%%%%%%%%%%
\newsec{Concluding remarks}
%%%%%%%%%%%%%%%%%%%%%%%%%%%%%%

In  this paper we  have argued that  naked bulk
singularities  of   gravitational backgrounds 
 dual to gauge theories can 
be resolved by introducing an analog 
of an IR cutoff in  gauge theories  into 
 the supergravity background.
 As a new 
explicit realization of this  proposal we demonstrated the 
resolution of the singularity of fractional D3-branes on conifold
background
 by the  compactification of the gauge-theory space-time 
   on $R\times S^3$ or $S^4$ with  
sufficiently large radius.

Unlike the original 
KT solution \KT, the resulting supergravity backgrounds discussed 
here are nonsupersymmetric.
 This should not be  too  surprising,
as our solutions are certain  deformations of the KT background
  which 
had only $\N=1$ supersymmetry in four dimensions.     
An interesting question is whether one can preserve supersymmetry 
in the process 
 compactification of  gauge theories with reduced supersymmetry, 
and what would be  their gravity duals.

 A promising starting 
point to address this question 
 is the so called $\N=2^*$ RG flow describing a
mass deformed $\N=4$ gauge theory. 
The corresponding supergravity solution 
 was found in \PW\ (PW), 
 and the 
realization of the gauge-gravity duality  in this case 
was explained in detail 
\refs{\bpp,\j}. It is  straightforward to 
construct a  linearized solution for  the gravitational background 
dual to the mass deformed $R\times S^3$  compactified $\N=4$ 
SYM theory. 
In fact, 
 the solution (and its  supersymmetries)
are precisely the same as in the original PW  construction. 
The physical explanation for this is that 
 the linearized solution 
effectively probes the UV dynamics of the gauge theory, where the 
compactification is actually irrelevant.

 A highly nontrivial 
question is whether one can find the  full nonlinear (supersymmetric?) 
solution
in this case. 
An intuitive reason   for why this  solution may exist is the following.
As explained in \bpp , the  PW flow is dual in the IR 
to a special vacuum point in  the $\N=2^*$ moduli space. Neither 
the gauge theory nor  gravity is able to explain what picks 
out this particular vacuum.
 The problem would be resolved if 
we assume the existence of an  analogous RG flow for the 
compactified $\N=4$ SYM theory. 
Indeed,  the adjoint  scalar 
coupling to the   curvature of $S^3$
  would lift all of the moduli space,  
apart from an isolated point; the 
conjecture is that the $\N=2^*$ vacuum 
of the PW flow is precisely the one 
surviving under the $S^3$ compactification. 
Finally, there is an interesting enhancon phenomenon in the 
PW geometry. The size of the $S^3$ in the ``compactified'' flow
produces a new mass scale in the geometry.
 One could imagine 
a phase transition originating from an  interplay between 
 the mass scale in the $\N=4$ 
deformation and the scale introduced by the
$S^3$.

\bigskip
%\noindent
{\bf Acknowledgements}
%%%%%%%%%%%%%%%%%%%%%%%%%%%%%%%

\noindent
We are grateful to Philip Argyres, 
Oliver DeWolfe, Sergey  Frolov, Gary Horowitz, Leo Pando Zayas, 
and Joe Polchinski    for
useful discussions. A.B. would like to thank 
the Aspen Center for Physics and the High Energy Particle 
Physics Theory Group at the University of Michigan for 
hospitality.
The work of A.B. is  supported in part 
by the NSF under grant No. PHY97-22022 and PHY99-07949. 
The work of 
 A.A.T.  is supported in part by the 
 DOE grant  DE-FG02-91ER-40690, 
as well as by the  PPARC SPG 00613,
 INTAS  99-1590 and  CRDF RPI-2108 grants.

%%%%%%%%%%%%%%%%%%%%%%%%%%%%%%%%%%%%%%%%%%%%%%%%%%%%%%%%
\vfill\eject
\listrefs
\end